\documentclass[a4paper,twocolumn,fleqn]{article} 

\usepackage{ist}
\usepackage{graphicx}
\usepackage{subcaption}
\usepackage[font=small,labelfont=bf]{caption}

\usepackage{booktabs} 
\usepackage{amsfonts} 
\usepackage{amsmath} 
\usepackage[numbers,sort&compress]{natbib}

\DeclareMathOperator*{\argmin}{argmin}
\DeclareMathOperator*{\argmax}{argmax}
\title{Language-based Color ISP Tuning}

\author{Owen Mayer~$^1$, Shohei Noguchi~$^2$, Alexander Berestov~$^1$, Jiro Takatori~$^2$ \\ $^1$Sony Corporation of America, San Jose, CA USA \\ $^2$Sony Corporation, Tokyo, Japan}
\date{} 

\begin{document} 

\maketitle 

\thispagestyle{empty} 

\begin{abstract}
	We propose a method for tuning the parameters of a color adjustment Image Signal Processor (ISP) algorithmic ``block" using language prompts. This enables the user to impart a particular visual style to the ISP-processed image simply by describing it through a text prompt. To do this, we first implement the ISP block in a differentiable manner. Then, we define an objective function using an off-the-shelf, pretrained vision-language-model (VLM) such that the objective is minimized when the ISP-processed-image is most visually similar to the input language prompt. Finally, we optimize the ISP parameters using gradient descent. Experimental results demonstrate tuning of ISP parameters with different language prompts, and compare the performance of different pretrained VLMs and optimization strategies.
\end{abstract}
\section{1. Introduction}
\label{sec:intro}


Modern digital cameras and camera phones utilize an ISP (Image Signal Processor) that processes the raw sensor data into an image that matches human-visual expectations. Typically, an ISP is comprised of a sequence of distinct algorithmic “blocks,” each of which handle a specific task within the ISP pipeline, such as Demosaicing, noise reduction, color adjustments, compression, etc~\cite{ramanath2005color,karaimer2016software}. While recent efforts have been made to replace on-device ISPs with neural networks~\cite{ignatov2020replacing, santos2025isp, conde2024nilut}, the typical modern ISP is implemented using rule-based and hand-crafted algorithms due to their power efficiency and consistency~\cite{yoshimura2024pqdynamicisp}.

The inner working of ISP blocks requires the setting of parameters, such as coefficients and thresholds, which critically impact the visual quality of the final processed image~\cite{karaimer2016software}. As a result, tuning the ISP parameters is a significant and important problem that affects the camera’s visual quality. Due to the difficulty of manually tuning ISP parameters, efforts have been made to develop automatic tuning algorithms~\cite{nishimura2018automatic, mosleh2020hardware, shi2022refactoring, kakarala2022cost, bianco2013color}. Furthermore, researchers have developed machine-learning-based ISP tuning methods~\cite{kim2020dnn, tseng2019hyperparameter, yang2020effective, tseng2022neural, kim2023learning, wang2024adaptiveisp, yoshimura2023dynamicisp}, which have recently shown great promise.

In this work, we investigate the automatic tuning of a color adjustment ISP block, and in particular the tuning of a $3\times 3$ color transformation matrix~\cite{bianco2013color}. In a typical ISP, the color adjustment block(s) perform two functions: 1) \textit{color correction}, i.e. correcting for spectral sensitivities of the imaging system and color-casts due to illumination~\cite{bianco2013color, kim2025ccmnet, shen2017color}, and 2) \textit{color enhancement}, i.e. imparting a visual style, feeling, or mood to the image using color~\cite{bonneel2013example, ke2023neural} referred to as \textit{color-grading}  in cinema~\cite{van2014color}. Color correction optimization is well studied, and typically uses calibrated images with known color values as the optimization target~\cite{bianco2013color, yamakabe2020tunable, yoshimura2024pqdynamicisp}. 

Color enhancement is inherently a more subjective problem, since the preferred visual style of an image changes from person-to-person. 
For example, a study in~\cite{bychkovsky2011learning} found that given the same input unenhanced image, different professional photographers added their own unique visual enhancements to it, exhibiting different style preferences. 
To ease the process of image enhancement, researchers have proposed approaches that use reference stylized images as targets for color enhancement optimization, which maps the visual style of the reference image(s) onto the input image. This allows the user to impart their preferred style by selecting example reference images. Several works achieve this by training a neural-network based ISP~\cite{ke2023neural, larchenko2025color}, filter estimation~\cite{bonneel2013example}, or ISP parameter optimization~\cite{tseng2022neural}.

\begin{figure}[!t]
	\centering
	\begin{subfigure}{0.98\columnwidth}
		\includegraphics[width=\linewidth]{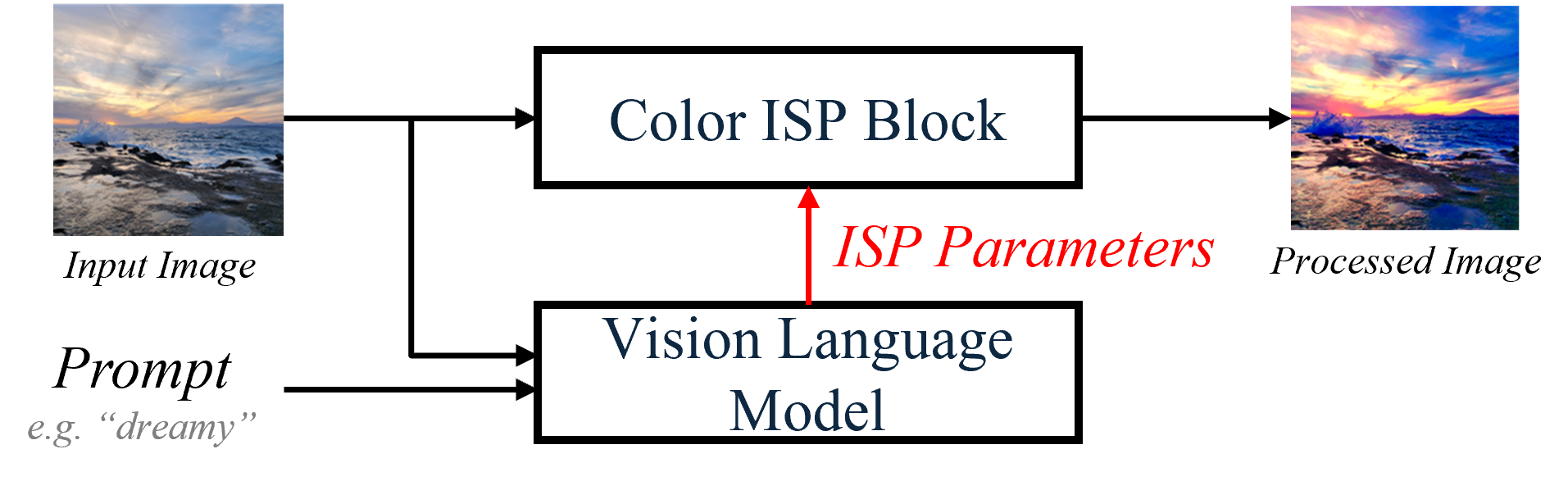}
	\end{subfigure}\vspace{1em}

	\hfill
	\begin{subfigure}[t]{0.19\columnwidth}
		\includegraphics[width=\linewidth]{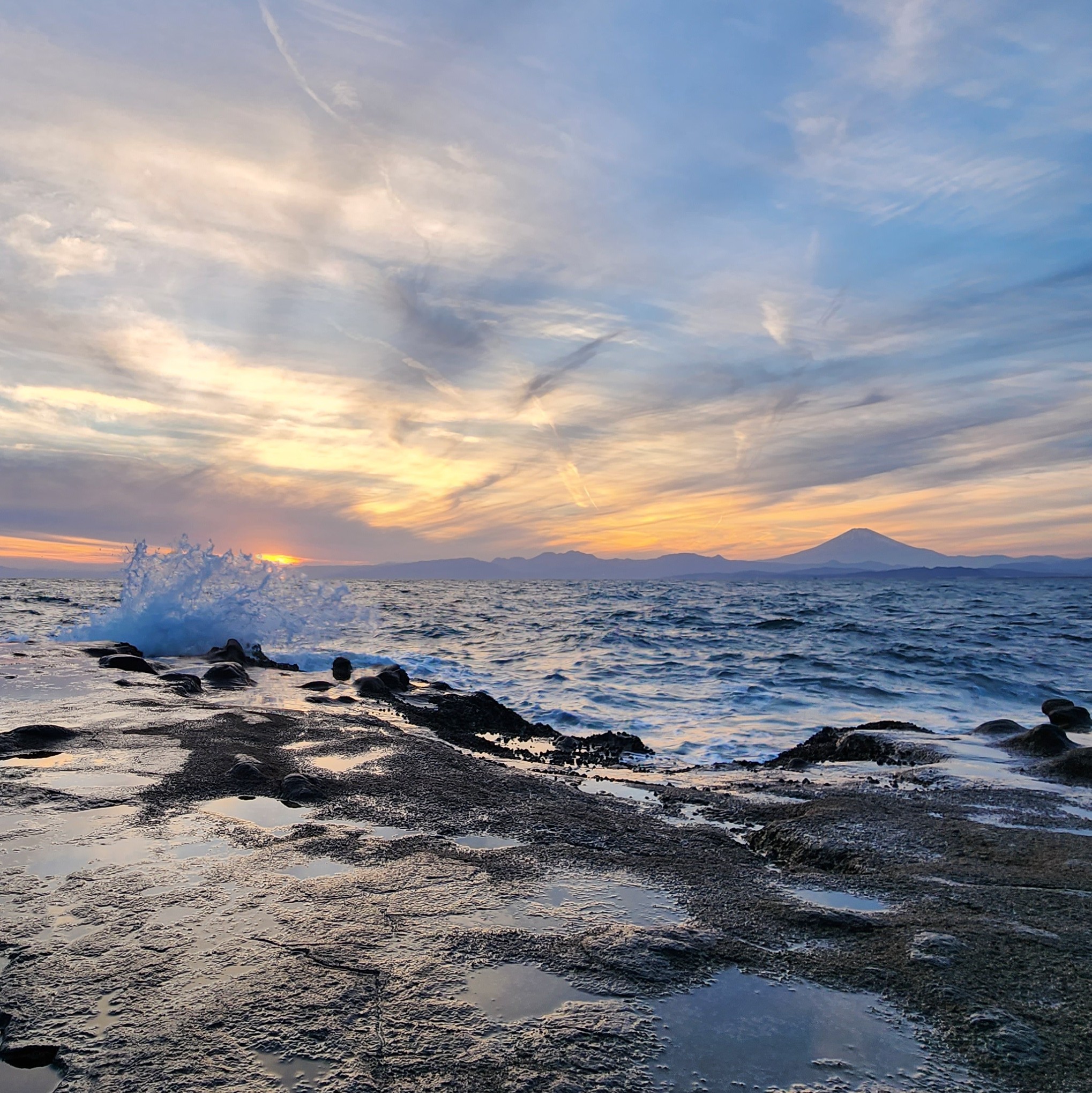}
		\caption{Input}
	\end{subfigure}\hfill
	\begin{subfigure}[t]{0.19\columnwidth}
		\includegraphics[width=\linewidth]{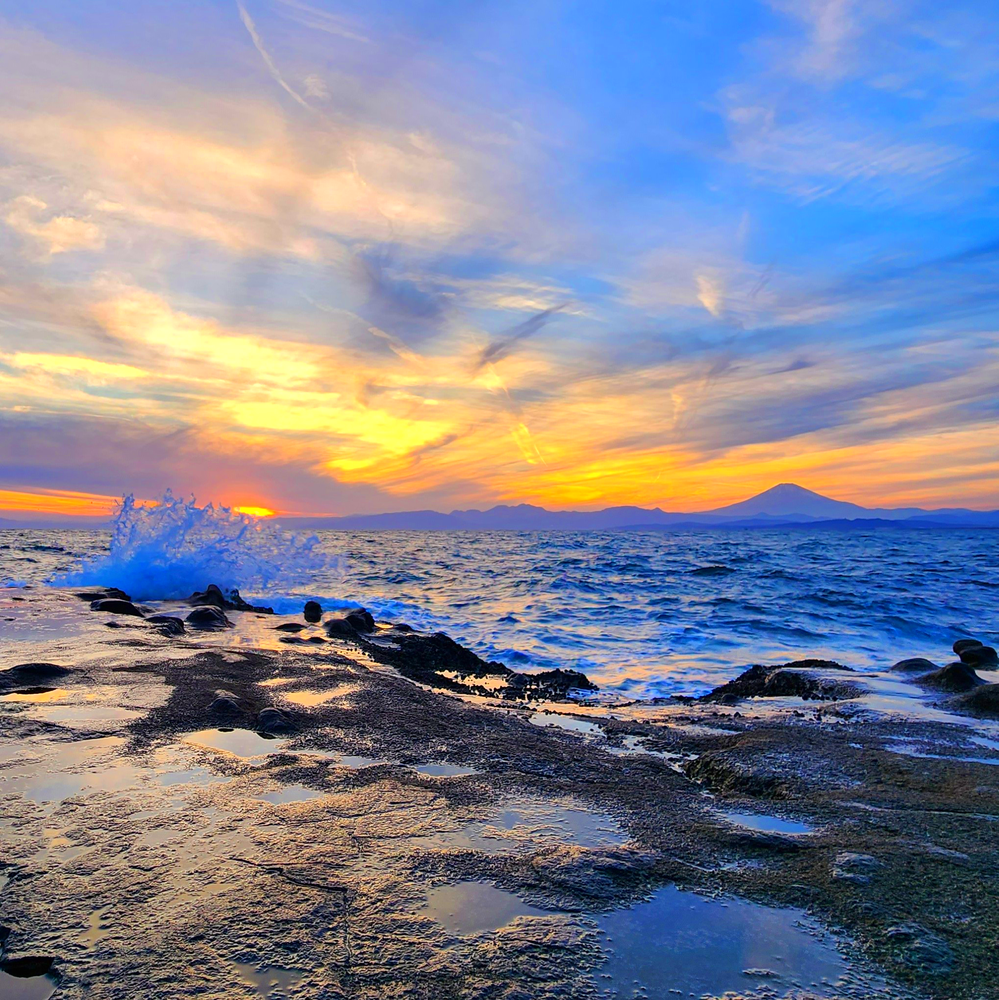}
		\caption{``Happy"}
	\end{subfigure}\hfill
	\begin{subfigure}[t]{0.19\columnwidth}
		\includegraphics[width=\linewidth]{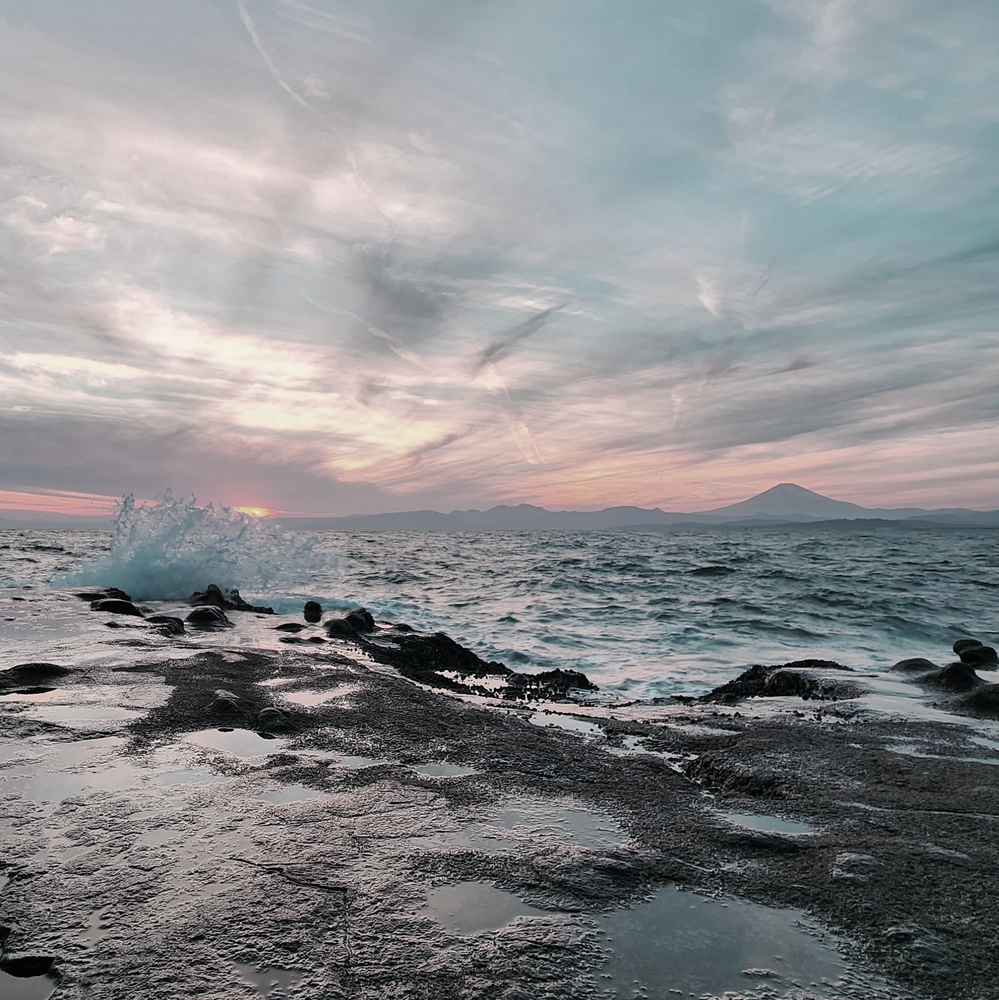}
		\caption{``Sad"}
	\end{subfigure}\hfill
	\begin{subfigure}[t]{0.19\columnwidth}
	\includegraphics[width=\linewidth]{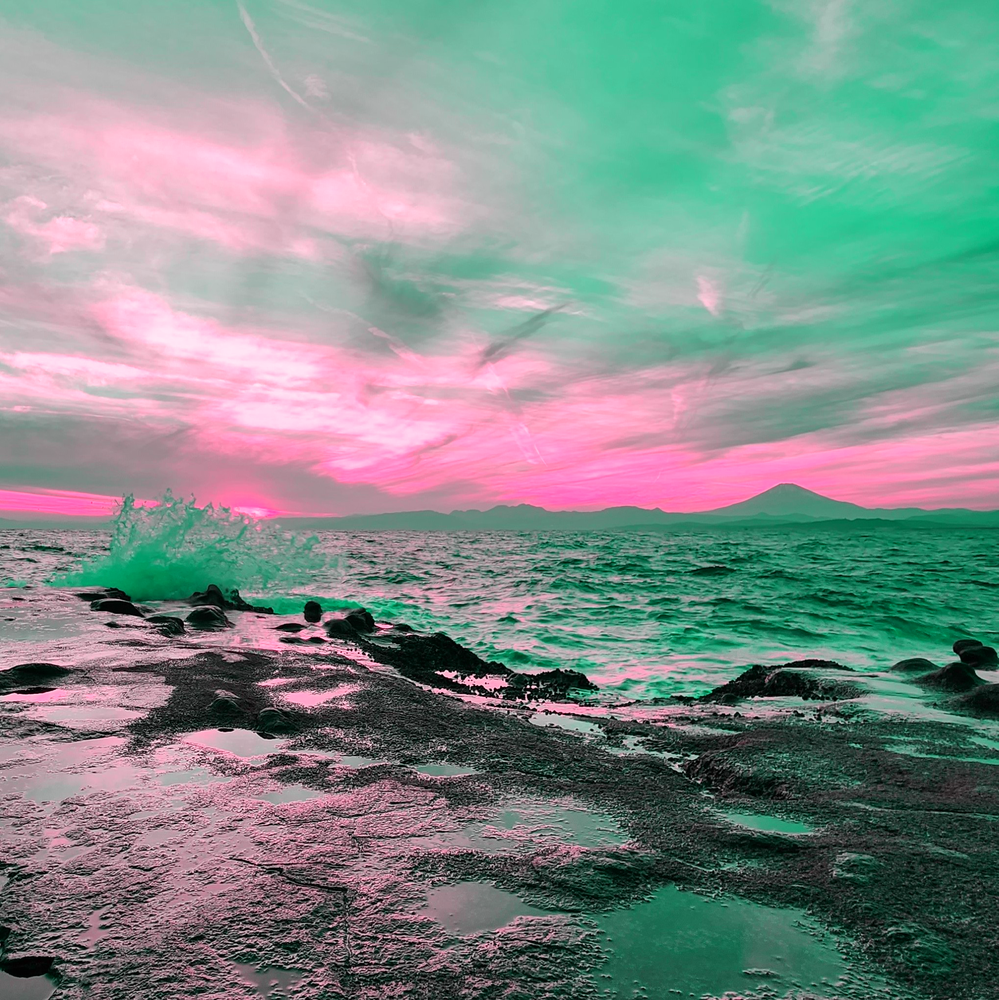}
	\caption{``Matrix Movie"}
\end{subfigure}\hfill
	\begin{subfigure}[t]{0.19\columnwidth}
	\includegraphics[width=\linewidth]{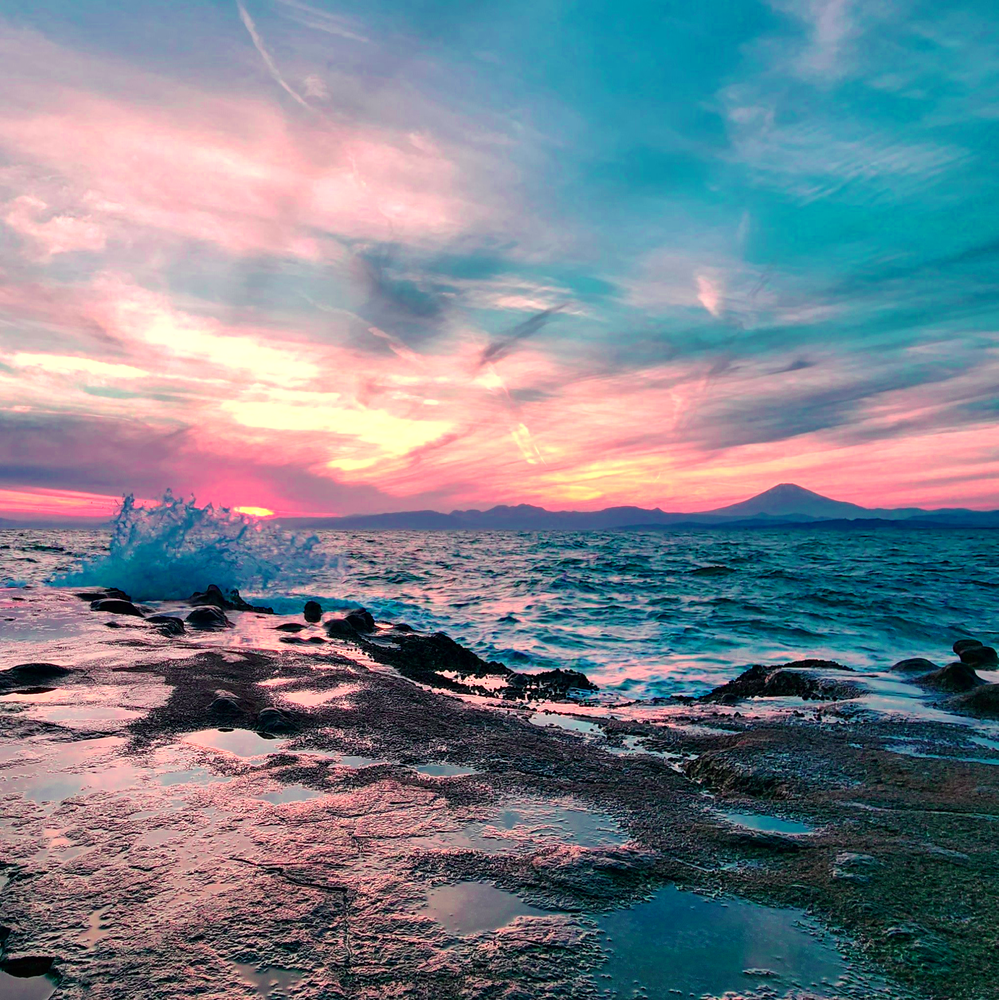}
	\caption{\scriptsize``Many Likes on Insta"}
\end{subfigure}\hfill
	\caption[Short Caption]{
	Our method tunes ISP parameters based on an input language prompt. The tuned parameters are then used to process the input image, resulting in a stylized image matching the prompt.
	}
	\label{fig:top}
\end{figure}

In this work, we take a different approach to color enhancement ISP tuning by proposing a method for optimizing a color adjustment ISP block that uses \textit{language prompts} as a style reference. To do this, we leverage recent advances in large vision language models (VLMs)~\cite{radford2021learning, ilharco_gabriel_2021_5143773, wang2022exploring, wang2022clip}, which connect the visual and language modalities.
We define an optimization objective that is minimized when the ISP parameters are tuned such that the output processed image is most similar to input text prompt, e.g. ``A vibrant photo." Similarity is computed by the Contrastive Language Image Pretraining (CLIP) VLM~\cite{radford2021learning}. Optimization is then performed by gradient descent directly on the ISP parameters. We note that no training of the VLM is required, and that off-the-shelf, pretrained models give sufficient results. Tuning of the ISP parameters is achieved in a few hundred optimization steps.

Our method provides a new and unique way to interact with and tune ISP parameters, and a novel way to impart visual styles to images: via language. See Fig.~\ref{fig:top} for examples. While other works have considered language-based image enhancement~\cite{fu2024mgie, nguyen2024instruction, chai2025giftcomesgoldpaper, li2024coco}, none use language for ISP tuning. Furthermore, in Sec. 2.3 we show that fully-neural-network-based enhancement methods such as~\cite{fu2024mgie} leave behind visual artifacts (``hallucinations") that our ISP tuning approach does not. To our knowledge, we are the first work on language-based ISP tuning, and the first work on the application of language-based color ISP block tuning. 
\begin{figure*}[!t]
	\centering
	\includegraphics[width=0.8\textwidth]{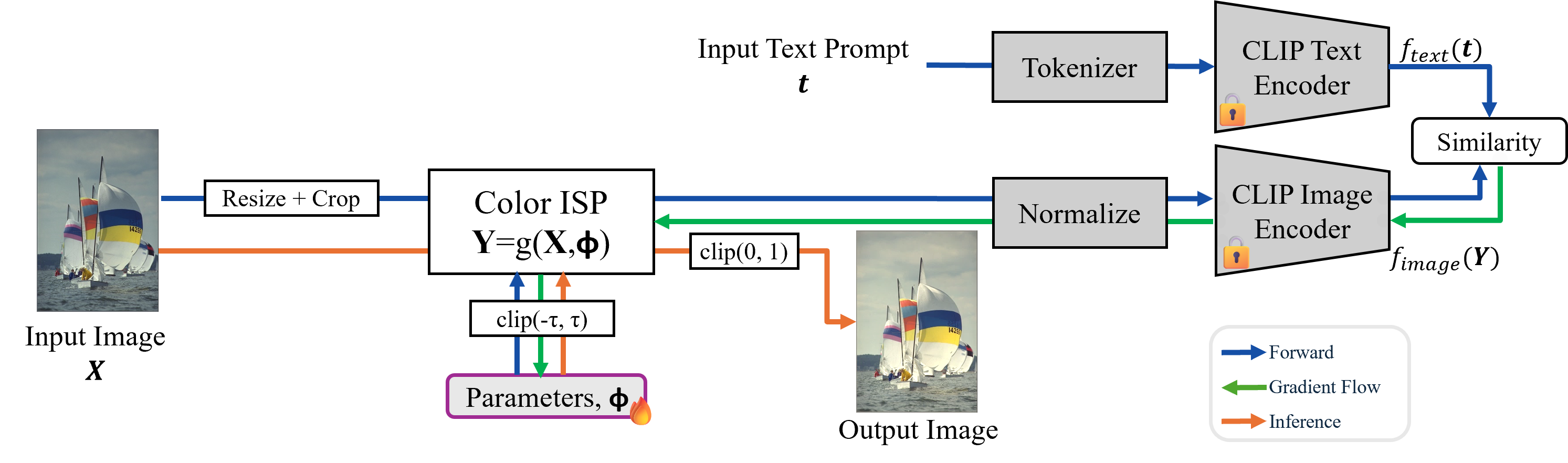}
	\caption{Our proposed language-based color-ISP-parameter optimization system. We use the CLIP vision-language model to define an objective that is minimized when the visual appearance of the input processed image is most similar to the description of the input language prompt. A gradient based solver is used to iteratively tune the ISP parameters.}
	\label{fig:system}
\end{figure*}

\section{2. Proposed ISP Tuning Method}

We propose a method to optimize the parameters, $\mathbf{\phi} \in \Phi$, of an ISP block. We first define an input RGB image $\mathbf{X} \in \mathbb{R}^{3 \times H \times W}$ with spatial dimensions $H$ and $W$ and 3 color channel dimensions \textit{R, G, B}, and an ISP block $g: \mathbb{R}^{3\times H \times W} \times \Phi \rightarrow \mathbb{R}^{3 \times H \times W}$ that maps an input image and ISP parameters to an output image  $\mathbf{Y} \in \mathbb{R}^{3 \times H \times W}$ of the same dimensions. Refer to Fig.~\ref{fig:system} for the full system diagram.

A color adjustment ISP block $g_{color}$ is utilized, comprised of a linear matrix multiplication; 
\begin{equation}
	g_{color}(\mathbf{X}, \phi) = \mathbf{M_\phi}\mathbf{X},
	\label{eq:gcolor}
\end{equation}
with $\mathbf{M_\phi}$ a 9 parameter $3\times3$ matrix 
\begin{equation}
\mathbf{M_\phi} = \begin{bmatrix} 1-\phi_{11} &  \phi_{12} & \phi_{13}\\ \phi_{21} & 1-\phi_{22} & \phi_{23}\\ \phi_{31} & \phi_{32} & 1-\phi_{33}, \end{bmatrix}.
\end{equation}
The 3x3 linear matrix formulation is a standard practice for color transformation, as found in~\cite{bianco2013color, kakarala2022cost, liu2024learnable, yoshimura2024pqdynamicisp, kim2025ccmnet}.

Parameters are initialized as $\phi_{ij} = 0~\forall~i,j$ and constrained such that $|\phi_{ij}|\leq \tau~\forall~i,j$.  Parameters are further constrained such that $ \left[1, 1, 1 \right]^T\mathbf{M}=\left[1, 1, 1 \right]^T$, where all rows sum to 1 ensuring that the white point is conserved~\cite{bianco2013color,kakarala2022cost}.

To enable use of a gradient-based solver, we require computation of $\partial g_{color} / \partial \phi_{i,j}$. Since our color ISP block is defined as a linear matrix multiplication in~\eqref{eq:gcolor}, the gradients are straightforward to compute by a standard deep learning framework.

\subsection{2.1 Language-based Objective}

The goal of optimization is to obtain parameters $\phi^*$, such that the processed image $\mathbf{Y}^*=g(\mathbf{X},\phi^*)$ has a visual appearance matching the language prompt description $\mathbf{t} \in \mathbb{T}$. Here $\mathbb{T}$ represents the space of text prompts, which is typically a short phrase in this work, e.g. $\mathbf{t}=$``A dreamy photo."

To do this, we utilize a pretrained CLIP (Contrastive Language-Image Pretraining) vision-language model~\cite{radford2021learning, ilharco_gabriel_2021_5143773, wang2022exploring}, which connects the visual and language modalities by computing similarity between an image and a text prompt. The CLIP model is comprised of two encoders; an image encoder 
$f_{image} : \mathbb{R}^{3\times H \times W} \rightarrow \mathbb{R}^F$, and text encoder
$f_{text} : \mathbb{T} \rightarrow \mathbb{R}^F$, which map an input image and input text prompt, respectively, into a common $F$-dimensional embedding space. The CLIP text encoder requires a tokenizer to first convert the text into a numerical representation, and is typically provided with existing pretrained CLIP models.

Next, a similarity measure 
$S : \mathbb{R}^F \times \mathbb{R}^F \rightarrow [0, 1]$ is defined between $f_{image}$ and $f_{text}$. 
We follow the original CLIP paper~\cite{radford2021learning} by using a cosine similarity,
\begin{equation}
	S(\mathbf{a},\mathbf{b}) = \frac{\mathbf{a} \cdot \mathbf{b}}{||\mathbf{a}|| ||\mathbf{b}||}
\end{equation}
however, other similarity measures can be explored~\cite{wang2022exploring,chou2024embedding}.

Finally, the optimization objective is defined,
\begin{equation}
	\phi^* = \argmax_\phi \; S\left[
	f_{image}\left(\mathbf{M_\phi}\mathbf{X}\right), 
	f_{text}\left(\mathbf{t}\right)
	\right],
	\label{eqn:optimization}
\end{equation}
which is solved by finding the $\phi$ such that the processed image is most visually similar to the input text prompt.

We further define a ``two-prompt'' optimization,
\begin{align}
	\phi^*_{A,B,\alpha} = \argmin_\phi \left\Vert \text{softmax} \left(
	s_{XA\phi}, s_{XB\phi}
	\right) - \alpha \right\Vert ^2
	\label{eqn:opt_2prompt}
\end{align}
where $s_{XA\phi}=S\left(
f_{image}\left(\mathbf{M_\phi}\mathbf{X}\right), 
f_{text}\left(\mathbf{t_A}\right)
\right)$. The two prompt optimization~\eqref{eqn:opt_2prompt}
tunes the ISP parameters to interpolate between text prompts $\mathbf{t}_A$ and $\mathbf{t}_{B}$, according to ratio $\alpha$, following the convention in~\cite{wang2022exploring}. Experimental results in Sec.~3.3 demonstrate the two-prompt objective enables fine-grain control over the visual appearance of the final processed image, and we recommend to the 2-prompt method as a best practice.

\subsection{2.2 Optimization}
Since the CLIP image encoder is neural network-based and the color ISP block is defined in a differentiable manner, the optimization objective is fully differentiable with respect to the ISP parameters. As a result, the optimization in Eq.\eqref{eqn:optimization} is solvable using gradient descent.

During optimization, the input image is resized, maintaining the original image aspect ratio, and center cropped to match the expected image size of the CLIP model. Further, the input to the image encoder is normalized according to constants derived by the training dataset, and are defined by OpenCLIP~\cite{ilharco_gabriel_2021_5143773}. The model ``VIT-B-32'' has feature dimension F=768 and expected image size 224$\times$224.
During inference, the output processed image is clipped to the range [0, 1] and converted to the uint8 datatype prior to display.

\begin{figure}[t]
	\centering
	\begin{minipage}[t]{0.95\linewidth}
		\includegraphics[width=\linewidth]{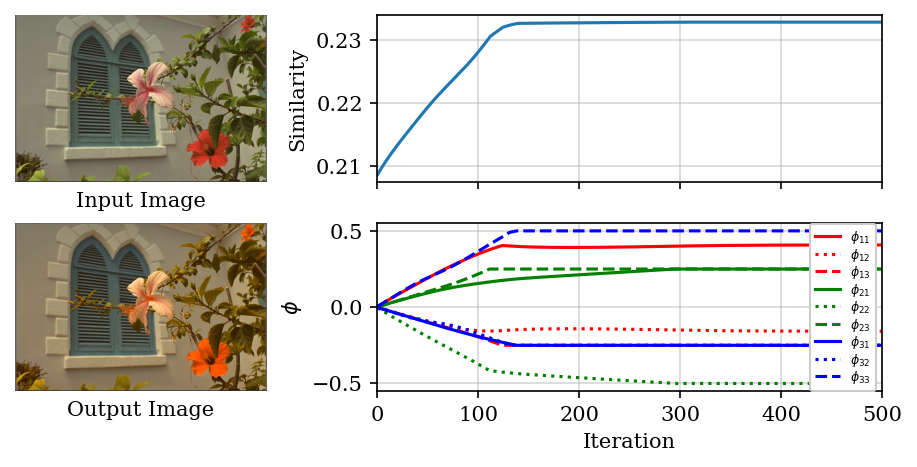}
	\end{minipage}
	\caption{Example parameter optimization with prompt $\mathbf{t}=$``A warm color palette photo." The color matrix parameters $\mathbf{\phi}$ are iteratively updated, resulting in a processed image with redder and warmer colors.}
	\label{fig:exp:opt_example}
\end{figure}

An example optimization is shown in Fig.~\ref{fig:exp:opt_example}, where the color ISP parameters are optimized using the prompt $\textbf{t} = $ ``A warm color-palette photo.'' During optimization we see that, in the lower right panel, the ISP parameters are adjusted, which then subsequently improves the similarity (upper right panel) between the prompt and processed image. The optimization is mostly converged within 100 iterations and fully in 400. Qualitatively, the final image exhibits a more ``warm''-like color palette, as observed by the overall reddened hues.

\subsection{2.3 Comparison to Neural Network-based Language Enhancement}

Existing methods for language-based image enhancement rely on neural networks to perform image processing~\cite{fu2024mgie, nguyen2024instruction}. While neural network approaches exhibit a high degree of expressiveness they have several drawbacks; they have strict resolution requirements and tend to impart visual artifacts (``hallucinations") to the output image. Our method does not suffer from such limitations.

\begin{figure}[t]
	\centering
	\hfill
	\begin{subfigure}[t]{0.36\columnwidth}
		\includegraphics[width=\linewidth]{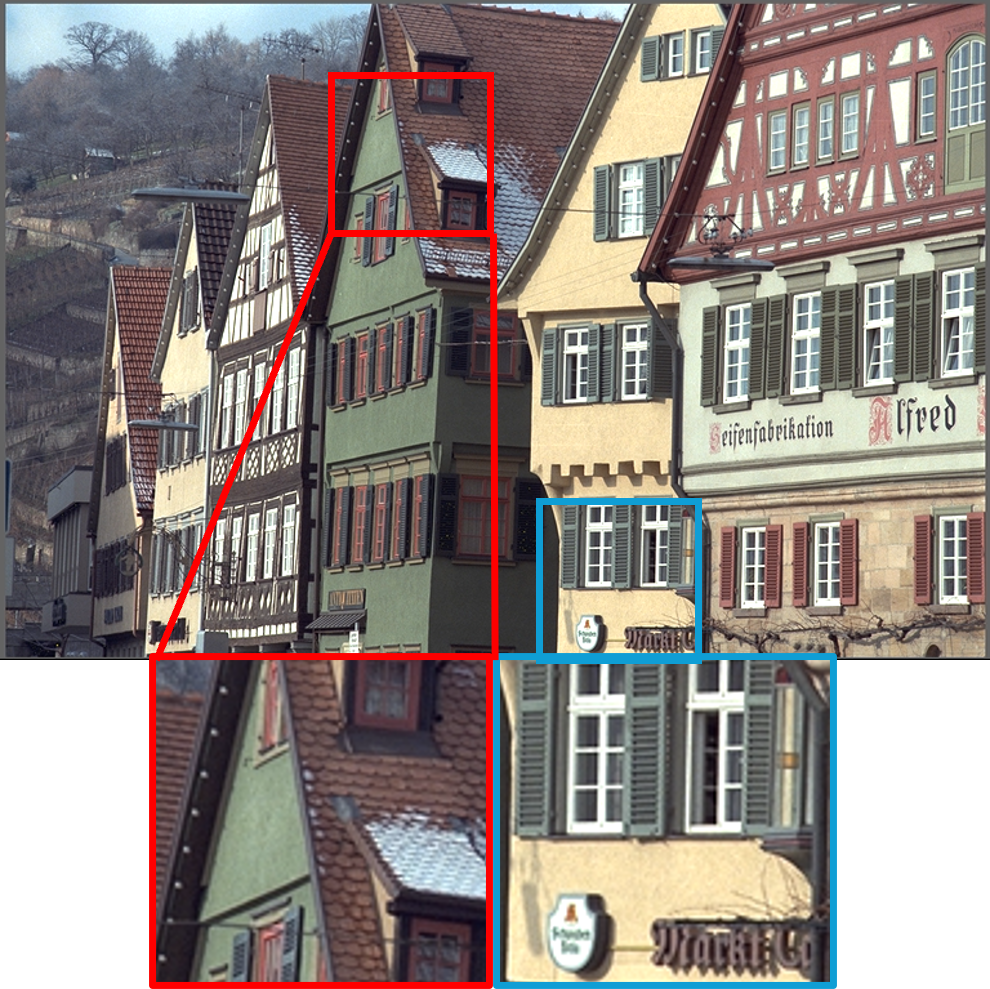}
		\caption{Input}
	\end{subfigure}\hfill
	\begin{subfigure}[t]{0.362\columnwidth}
		\includegraphics[width=\linewidth]{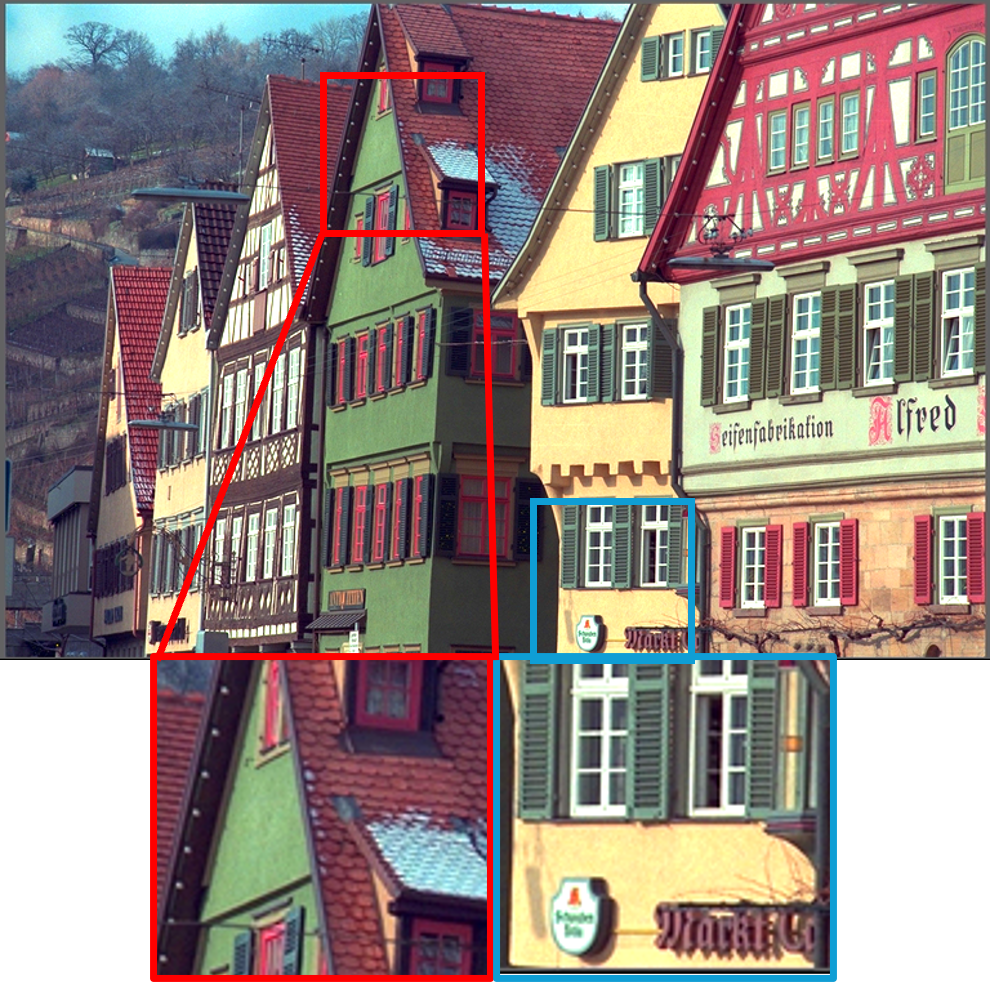}
		\caption{Our ISP Tuning}
	\end{subfigure}\hfill
	\begin{subfigure}[t]{0.252\columnwidth}
		\includegraphics[width=\linewidth]{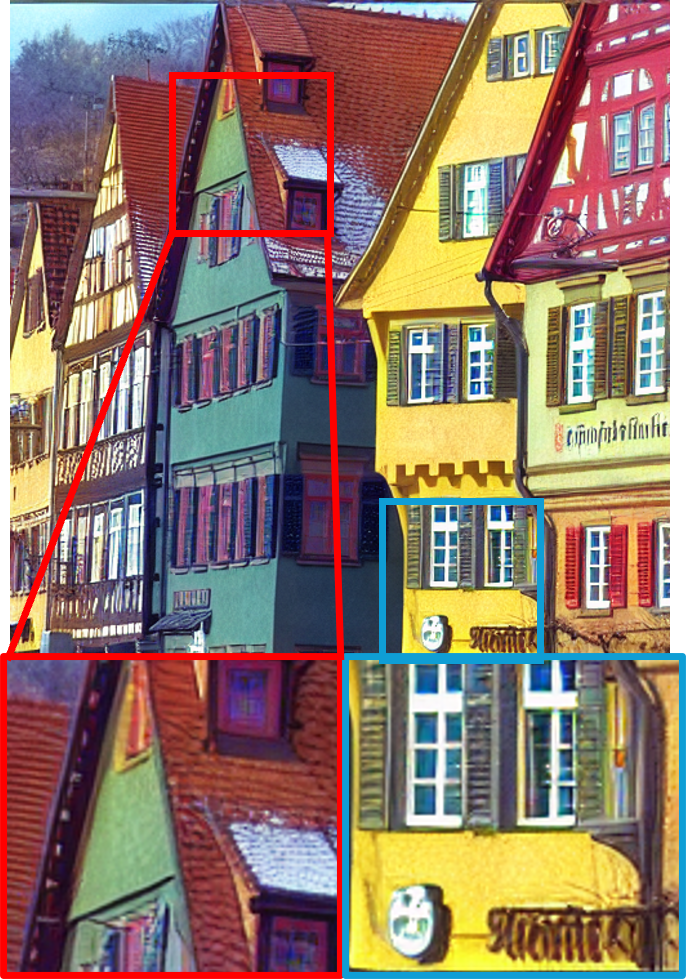}
		\caption{MGIE~\cite{fu2024mgie}}
	\end{subfigure}\hfill
	\caption[Short Caption]{
		Output processed images for our ISP tuning method and the MGIE neural-network based image enhancement method, using prompt ``A vibrant photo." Our method does not have strict resolution requirements as MGIE, and does not impart visual artifacts and hallucinations as shown in the zoomed in sections.
	}
	\label{fig:mgie}
\end{figure}

We demonstrate this by color-enhancing an image using the prompt ``A vibrant photo." using our method and the method in~\cite{fu2024mgie}, a state of the art language-based image editing and enhancement algorithm ``MGIE." The output processed images are shown in Fig.~\ref{fig:mgie}. First we note that the MGIE method imparts severe visual artifacts and hallucinations, as shown in the zoomed in regions. The textures of the roof are significantly changed (red inset), and the sign content is altered (blue inset). Second, we note that MGIE requires the input image to size $512 \times 512$ and so their result is center-cropped. ISP-tuning approaches do not suffer from these drawbacks. 

While the MGIE output is more vibrant looking, we note that the optimal ``vibrant photo" is subjective. We show in Sec.~3.2 and Sec.~3.3.D that adjusting the parameter clipping constraints and using the two-prompt tuning method overcomes this limitation by enabling control over the degree of visual style.

\section{3. Experimental Results}

We evaluated our approach by conducting a series of experiments, starting with 24 RGB images from the ``Kodak Lossless True Color Image Suite"\footnote{https://r0k.us/graphics/kodak/} as input images. 
Unless otherwise specified, we performed optimization for 1000 iterations using the Adam optimizer with learning rate$=2e-3$, parameter clipping level $\tau=0.25$, and CLIP model ``VIT-B-32'' with pretrained weights ``LAION-2B''~\cite{schuhmann2022laionb} implemented by OpenClip~\cite{ilharco_gabriel_2021_5143773}. Our method was implemented in the PyTorch framework~\cite{pytorch}, and ran on a single RTX3090 GPU. For a single image and prompt, optimization completed in approximately 30 seconds.

\subsection{3.1 Prompt variety}
Our method relies on the pretrained CLIP model to correctly interpret the 1) visual style aspects of the language prompt and 2) visual style aspects of the image, while being agnostic to the semantic content of the image. Recent works have suggested this to be possible, but also bring up several deficiencies~\cite{wang2022exploring, arias2025color, chai2025giftcomesgoldpaper}.


\begin{figure*}[t]
	
	\begin{minipage}[h]{\linewidth}
		~\hfill
		\begin{subfigure}[t]{0.9\linewidth}
			\includegraphics[width=\linewidth]{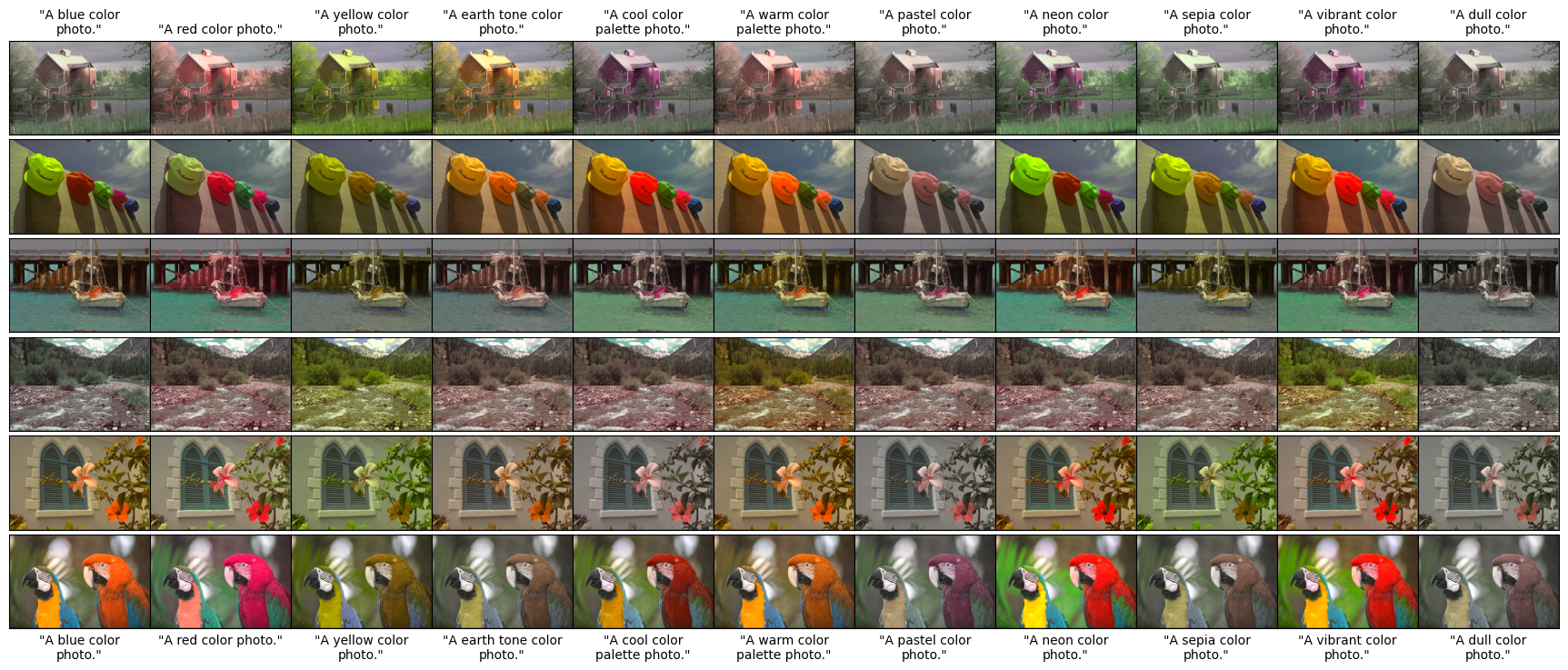}
			\caption{Color Prompts}
		\end{subfigure}
	\hfill~
	\end{minipage}
	
	\begin{minipage}[h]{\linewidth}
		~\hfill
		\begin{subfigure}[t]{0.42\linewidth}
			\includegraphics[width=\linewidth]{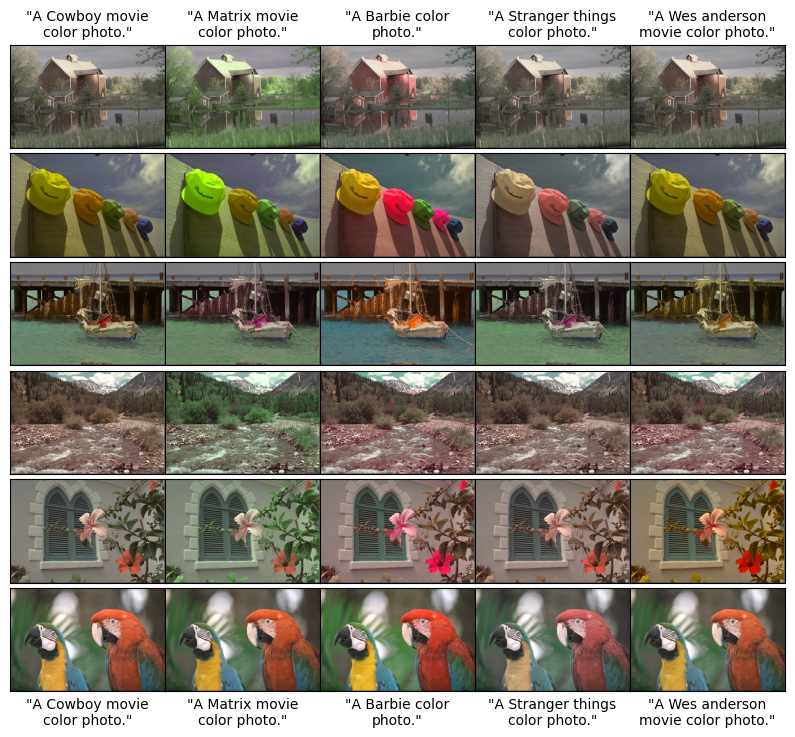}
			\caption{Cultural Prompts}
		\end{subfigure}
	\hfill
	\begin{subfigure}[t]{0.42\linewidth}
		\includegraphics[width=\linewidth]{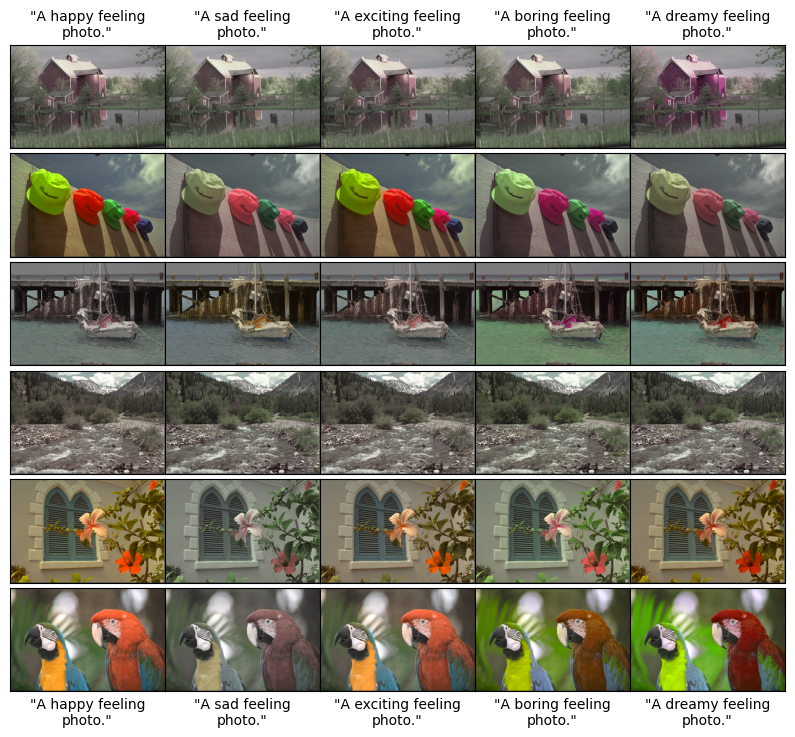}
		\caption{Abstract, Emotional Prompts}
	\end{subfigure}
	\hfill~
	\end{minipage}

	\caption{(Best viewed in color and zoomed.) Images processed after ISP tuning on varied language prompts, including explicit color descriptors, abstract emotions, and cultural references. Different visual styles can be achieved simply by describing it through a text prompt.}
	\label{fig:exp:target_examples}
\end{figure*}

\begin{figure*}[h!]
	\centering
	\begin{minipage}[t]{0.99\linewidth}
		\includegraphics[width=\linewidth]{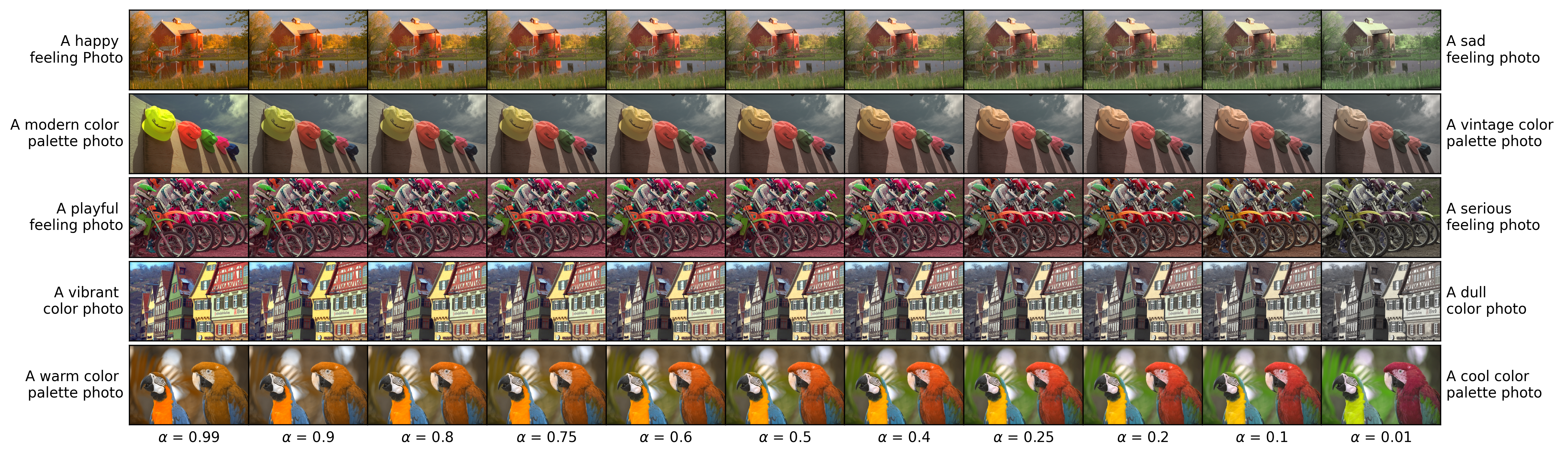}
	\end{minipage}\hfill%
	\caption{(Best viewed in color and zoomed.) Processed images after tuning using the ``2 prompt approach", which interpolates between two prompts according to parameter~$\alpha$. Fine grain control is achieved over the look and feel of the final processed image.}
	\label{fig:exp:2prompt}
\end{figure*}
To test this, we qualitatively evaluated how well our CLIP-based method performs with different target prompts and category of prompt. We evaluated with in the following categories: Explicit color descriptors (e.g. ``dark yellow," ``cool"), cultural reference (e.g. ``Matrix movie," ``Cowboy western"), and abstract/emotional (e.g. ``happy," ``dreamy").
Optimization was performed on all 24 images of the dataset with input prompt $\mathbf{t}=$ ``A \{prompt\} photo." Visual results are shown in Fig.~\ref{fig:exp:target_examples}, which show the output processed images. We found that the model works well for explicit color descriptors, for which there are commonly understood meanings. For example, ``warm" tuning results in processed images that have redder and yellower tones, whereas ``cool" tuning results in blue and purple tones. 

Interestingly, the tuning method was able to capture colors related to cultural references, such as the neon greens and purples of the ``Matrix movie" or the muted yellows of a country western ``Cowboy" movie, indicating that the vision language model we used was able to correctly encode these concepts.

Prompts that use abstract emotions did not work as well. For example the 1st and 4th row of images did not show much stylistic change from prompt-to-prompt. The other rows did show some change, but it was not clear that the final images are consistently well matched to the expected look. Still, we found that the tuning on abstract emotions were able to impart some visual qualities that, at least to these authors, were aligned with prompts. For example, ``happy" tuning resulted in brighter and more vibrant colors and ``sad" tuning resulted in dull and more muted colors. 

These results demonstrate the proposed tuning method can achieve visual styles matching the input language prompt. However, some prompts and images did not work as well. We attribute the instance of poor performance to inability of VLM to properly encode the stylistic meaning in the visual and/or language domain, and improvements might be made by finetuning the CLIP model to this task. 

\subsection{3.2 Two-prompt tuning}
A drawback of using language is that language itself can be ambiguous, yielding unexpected results. For example, the general concept of a ``warm" image may have a rough common meaning, but the exact appearance and optimal degree of ``warmness" may very from person-to-person. As a result finely controlling the ISP is important in the context of language tuning. The two prompt approach defined in~\eqref{eqn:opt_2prompt} addresses this by using a 2nd prompt to enforce additional opposite meaning (e.g. a ``warm" image \textit{and} not a ``cool" image) as well as enabling finer control in degree of the imparted visual style via interpolation parameter $\alpha$. In this experiment, we evaluated the two prompt approach.

To do this, we performed optimization using~\eqref{eqn:opt_2prompt}, with a range of 
$\alpha$
and 5 different pairs of prompts, shown in Fig.~\ref{fig:exp:2prompt}.

Each of the 5 prompt-pairs exhibit a visual style gradient determined by $\alpha$. For example, the ``happy feeling" ($\alpha=0.99$) photo shows bright, vibrant colors, whereas the ``sad feeling" ($\alpha=0.01$) photo shows duller, muted, and cooler colors. Importantly, \textit{the degree} of this visual appearance is smoothly interpolated, via the $\alpha$ parameter, demonstrating fine grained visual appearance control. Overall, we found this 2 prompt approach to give better results.

\subsection{3.3 Quantitative result and ablations}
\begin{figure}[t]
	\centering
	\hfill
	\begin{minipage}[h]{0.44\columnwidth}
			\hfill
			\begin{subfigure}[t]{0.5\linewidth}
					\includegraphics[width=\linewidth]{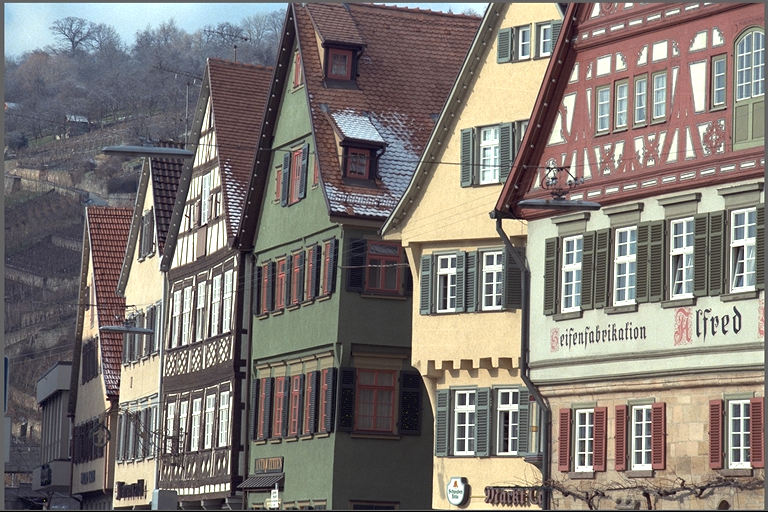}
					\caption{Input}
				\end{subfigure}\hfill~
			\\
			\begin{subfigure}[t]{0.45\linewidth}
					\includegraphics[width=\linewidth]{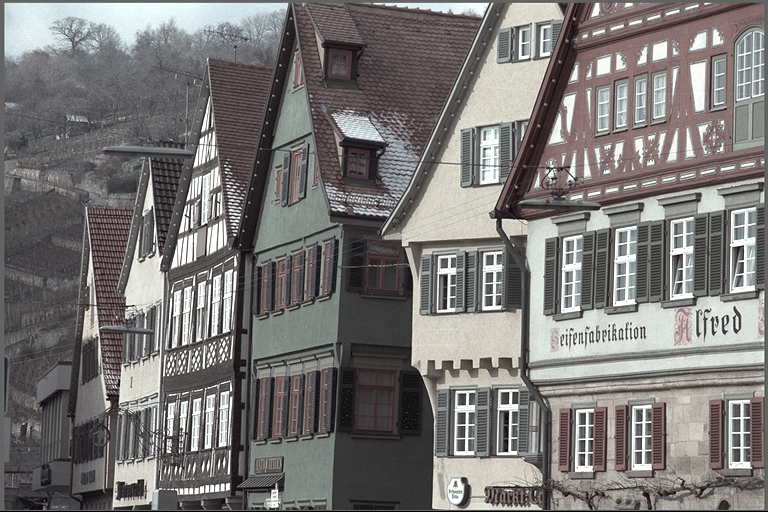}
					\caption{``Dull"  }
				\end{subfigure}
			\begin{subfigure}[t]{0.45\linewidth}
				\includegraphics[width=\linewidth]{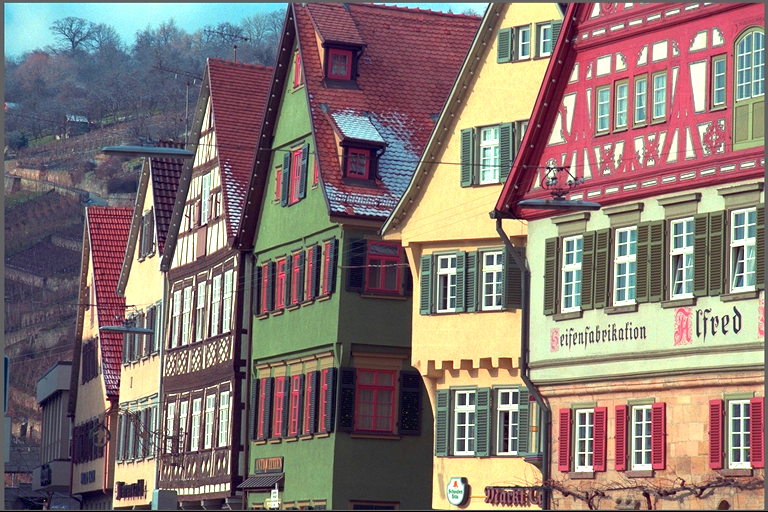}
				\caption{``Vibrant"}
			\end{subfigure}
		\end{minipage}\hfill%
	\hfill
	\begin{minipage}[h]{0.55\columnwidth}
			\begin{subfigure}[t]{\linewidth}
					\includegraphics[width=0.9\linewidth]{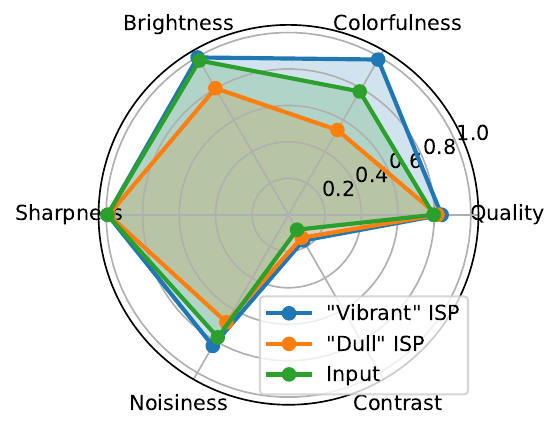}\vspace{-0.75em}
					\caption{CLIP-IQA Ratings}
				\end{subfigure}
	\end{minipage}\hfill%
	\caption{Processed images from tuning with ``vibrant" and ``dull" as input prompts. The CLIP-IQA "colorfulness" rating (d) shows significant differences between the images.}
	\label{fig:exp:colorfulness}
\end{figure}

To evaluate our design choices, we conducted an experiment providing a quantitative result. To do this, we compared the processed images from prompts ``A vibrant image" $\mathbf{Y*}_{v}$ and ``A dull image" $\mathbf{Y*}_{d}$ under different settings. Then, we evaluated the vibrant-dull image pairs by computing the 1) difference in ``Colorfulness`` CLIP-IQA score~\cite{wang2022exploring}, and 2) the difference in colorfulness metric~\cite{hasler2003measuring}. These values are reported in Table~\ref{tab:clipiqa_quant} as $\Delta$ \textbf{CLIP-IQA} and $\Delta \mathbf{C}$, where higher values indicate the method was able to impart a higher difference in colorfulness. An example is shown in  Fig.~\ref{fig:exp:colorfulness},  where the bottom right panel shows the CLIP-IQA scores, which quantifies the ``colorfulness" in the input and processed images. 

We evaluated our method's performance for different 1. prompt constructions, 2. pretrained CLIP models, 3. solvers, and 4. parameter clipping level $\tau$. 
\subsubsection{3.3.A Prompt Construction}
The different prompt constructions are shown in Table~\ref{tab:prompts}. We found \textbf{B} ``A \{prompt\} photo" to performed best, whereas prompts ``\{prompt\}" and ``A \{prompt\} photo of \{content description\}" performed the worst. This finding aligns with other works~\cite{wang2022exploring} that found the choice of prompt impacts the performance of CLIP models. Interestingly, adding content descriptions with prompt D did not improve the performance, and was possibly causing the optimization objective to focus on the content, rather than introducing extra context as was intended.
\subsubsection{3.3.B Model Choice}
Of the pretrained models available in OpenCLIP~\cite{ilharco_gabriel_2021_5143773}, we found architecture \textit{ViT-B/32} with pretrained weights \textit{laion2b\_s34b\_b79k} to perform the best. Using pretrained weights by OpenAI or using a larger model \textit{ViT-L/14} performed worse. We found it interesting that the larger model performed worse. One possible explanation is a gradient issue encountered by using a larger model, which might be solved by further tuning the optimization hyperparameters but is outside the scope of this work.
\subsubsection{3.3.C Optimizer Choice}
We found that the choice of optimizer did not impact the result, with Adam, AdamW, and SGD performing nearly equally. SGD did give a slightly better CLIP-IQA score, but the difference was not significant. For each optimizer, a learning rate was empirically chosen that yielded best performance.

\begin{table}[t]
	\centering
	\vspace{1em}
		\caption{Quantitative results and system design evaluation}
	\scriptsize
	\begin{tabular}{ccccrr} \toprule
		\textbf{Prompt} & \textbf{Model} & $\mathbf{\tau}$ & \textbf{Optim.} &  $\Delta$\textbf{\tiny CLIP-IQA $\uparrow$} & $\Delta \mathbf{C} \uparrow$ \\ \midrule
		A  & ViT-B-32 / laion2b & 0.25 & adam & 0.24 & 29.9 \\ 
		\textbf{B}  & \textbf{ViT-B-32 / laion2b} & \textbf{0.25} & \textbf{adam} & \textbf{0.34} & \textbf{35.1} \\
		C  & ViT-B-32 / laion2b & 0.25 & adam & 0.33 & 32.6 \\
		D  & ViT-B-32 / laion2b & 0.25  & adam & 0.30 & 23.0\\ \midrule
		B  & ViT-B-32 / openai & 0.25  & adam & 0.20 & 24.2 \\
		B  & ViT-L/14 / laion2b & 0.25 & adam & 0.14 & 13.9\\ \midrule
		B  & ViT-B-32 / laion2b & 0.25 & adamw & 0.34 & 35.1\\
		B  & ViT-B-32 / laion2b & 0.25 & sgd & \textbf{0.35} & 35.1\\ \midrule
		B  & ViT-B-32 / laion2b & 0.33 & adam & 0.44 & 42.1\\
		B  & ViT-B-32 / laion2b & 0.50 & adam & 0.49 & 49.5\\
		B  & ViT-B-32 / laion2b & 1.00 & adam & 0.48 & 61.3\\ \bottomrule
	\end{tabular}
\label{tab:clipiqa_quant}
\end{table}

\subsubsection{3.3.D Parameter Clipping}
\begin{figure}[!t]
	\hfill
	\begin{minipage}[h]{0.24\columnwidth}
		~\hfill
			\begin{subfigure}[t]{0.45\linewidth}
					\includegraphics[width=\linewidth]{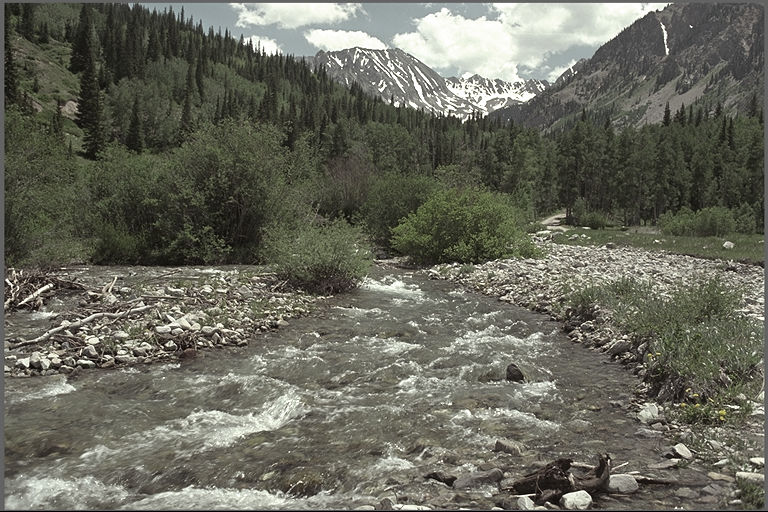}
			\end{subfigure}\hfill
			\begin{subfigure}[t]{0.45\linewidth}
			\includegraphics[width=\linewidth]{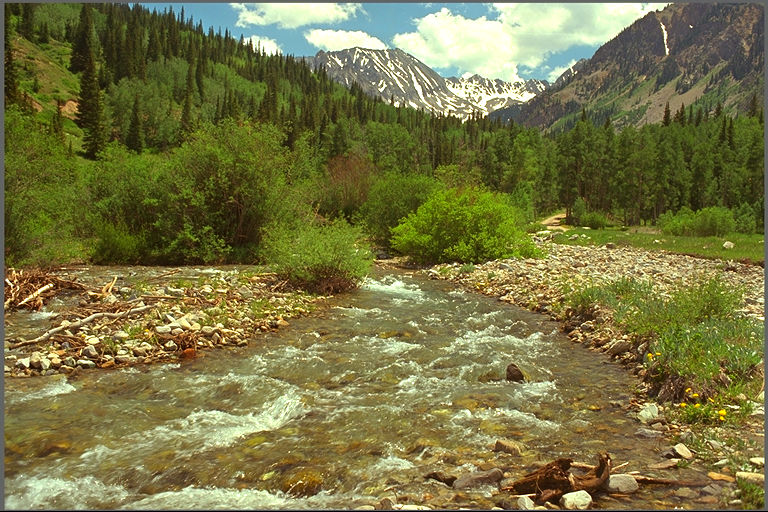}
		\end{subfigure}\hfill~
	\centering \small
	$\tau=0.25$
	\end{minipage}
	\hfill
	\begin{minipage}[h]{0.24\columnwidth}
		~\hfill
		\begin{subfigure}[t]{0.45\linewidth}
			\includegraphics[width=\linewidth]{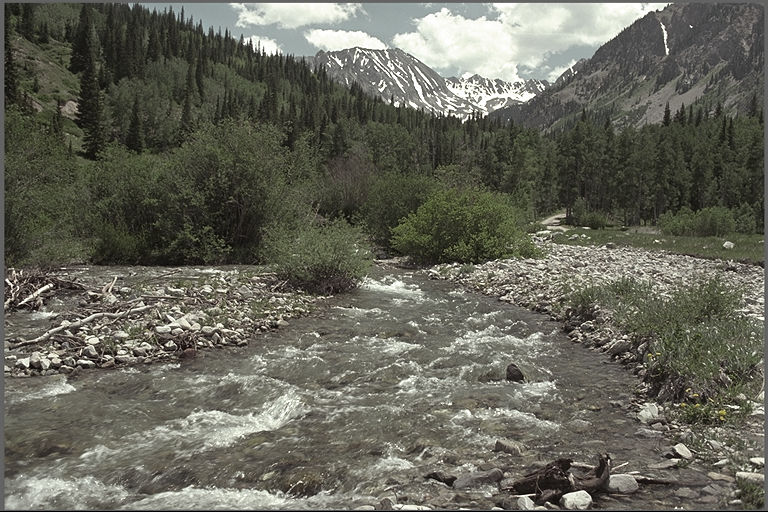}
		\end{subfigure}\hfill
		\begin{subfigure}[t]{0.45\linewidth}
			\includegraphics[width=\linewidth]{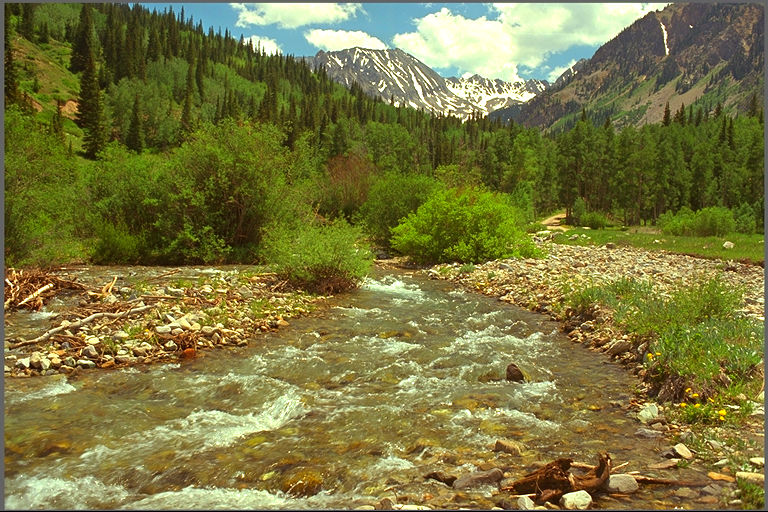}
		\end{subfigure}\hfill~
		\centering \small
		$\tau=0.25$
	\end{minipage}
	\hfill
	\begin{minipage}[h]{0.24\columnwidth}
		~\hfill
		\begin{subfigure}[t]{0.45\linewidth}
			\includegraphics[width=\linewidth]{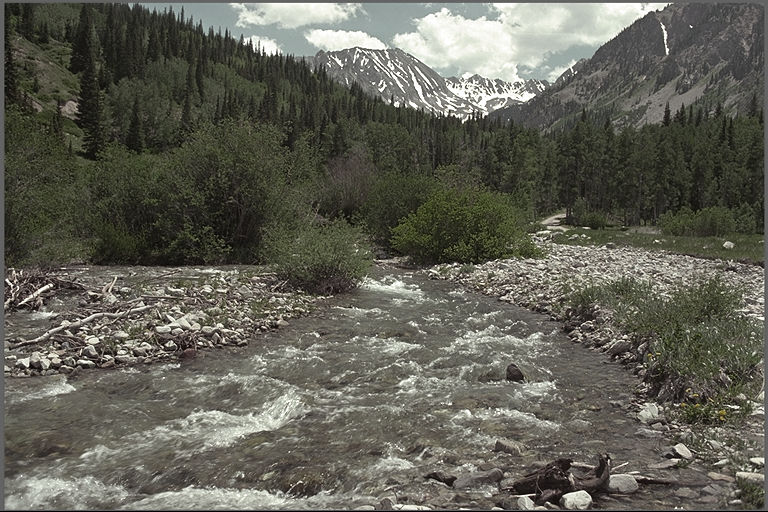}
		\end{subfigure}\hfill
		\begin{subfigure}[t]{0.45\linewidth}
			\includegraphics[width=\linewidth]{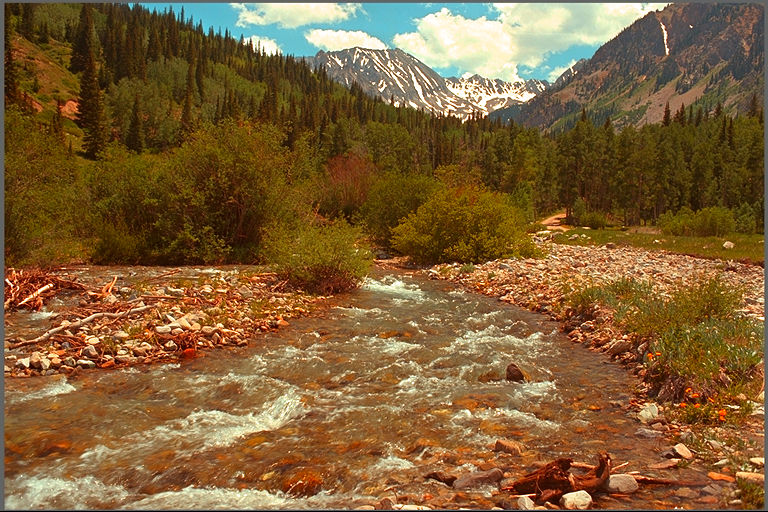}
		\end{subfigure}\hfill~
		\centering \small
		$\tau=0.25$
	\end{minipage}
	\hfill
	\begin{minipage}[h]{0.24\columnwidth}
		~\hfill
		\begin{subfigure}[t]{0.45\linewidth}
			\includegraphics[width=\linewidth]{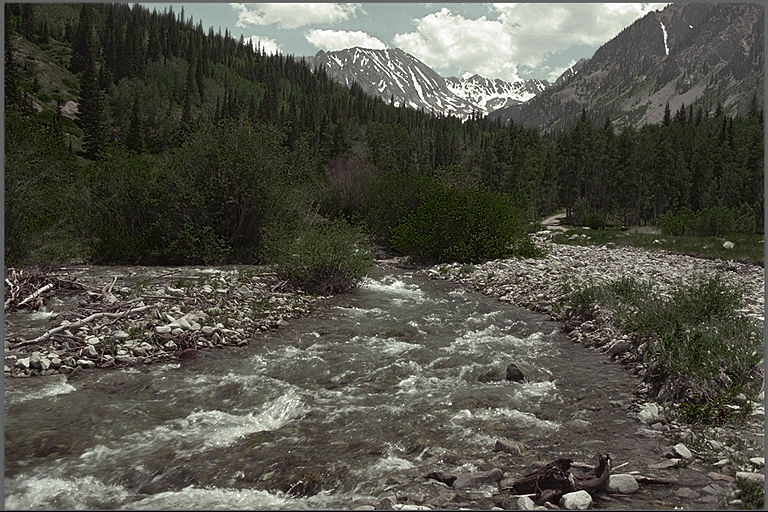}
		\end{subfigure}\hfill
		\begin{subfigure}[t]{0.45\linewidth}
			\includegraphics[width=\linewidth]{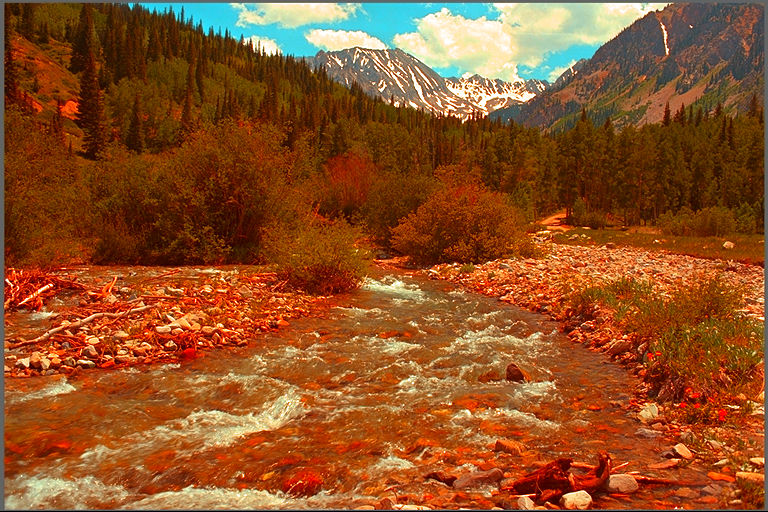}
		\end{subfigure}\hfill~
		\centering \small
		$\tau=0.25$
	\end{minipage}
	\hfill

	\caption[Short Caption]{
		Processed versions of \textit{Kodak Image 13} tuned with the prompts ``A dull photo" (left), and ``A vibrant photo" (right), at different coefficient clipping levels, $\tau$, of 0.25 to 1.0.
	}
	\label{fig:clipping}
\end{figure}

We tuned with different parameter clipping levels $\tau~\in~\{0.25, 0.33, 0.5, 1.0\}$, and found that relaxing the clipping level (larger $\tau$) resulted in greater colorfulness differences between the ``vibrant" and ``dull" processed images. Examples of tuning with different $\tau$ is shown in Fig.~\ref{fig:clipping}. While relaxing the clipping constraint enables the tuning to impart more significant color changes, we observed that $\tau \geq 0.5$ resulted in a unnatural looking processed images and so we empirically set $\tau = 0.25$.

\begin{table}[t]
		\vspace{1em}
	\caption{Different prompt constructions used in Experiment 3.2}
	\scriptsize
	\begin{tabular}{cl} \toprule
		\textbf{Prompt ID} & \textbf{Prompt Construction}  \\ \midrule
		A  & ``\{prompt\}" \\ 
		B  & ``A \{prompt\} photo"\\
		C  & ``A photo that appears \{prompt\}" \\
		D  & ``A \{prompt\} photo of \{content description\}" \\  \bottomrule
	\end{tabular}

	\label{tab:prompts}
\end{table} 

\section{4. Limitations}
Here, we discuss several limitations:
First, our tuning method uses a single image input, resulting in ISP parameters that are tuned for that specific image. Tuning per-image may be inefficient to deploy in practice, and so methods that tune global parameters applied to any image can be explored. Second, the range visual appearance is limited by the expressive power of the ISP block itself. Other ISP blocks can be explored for tuning, such as LUTs, however our method requires the ISP block to be defined in a differentiable manner. Third, using language to describe visual style is inherently subjective, which makes evaluation difficult. Evaluation may be benefited by the development of better metrics or by large-scale user surveys. Finally, the vision language model we used was trained on images and text that largely focused on the semantic content of the image, rather than stylistic elements, which may be causing worse performance. The method can be improved by finetuning the CLIP model on data labeled according to image stylistic elements. We hope that the idea of language-based ISP tuning can be improved upon to address these limitations.

\section{5. Conclusion}
We proposed a method to tune the parameters of a color ISP block, simply by describing the desired style via a language prompt. To do this, we used the CLIP vision language model to define an objective function on the target prompt and the processed image, and then performed gradient descent onto the ISP parameters themselves. Our method demonstrates a novel way to tune ISP parameters, and experimental results highlighted creative visual styles that can be achieved with this method, along with limitations.





\small
\bibliographystyle{IEEEtran}
\setlength{\bibsep}{2pt}
\bibliography{clip.bib, isp.bib, color.bib, misc.bib}

\begin{thebibliography}{10}
\providecommand{\url}[1]{#1}
\csname url@samestyle\endcsname
\providecommand{\newblock}{\relax}
\providecommand{\bibinfo}[2]{#2}
\providecommand{\BIBentrySTDinterwordspacing}{\spaceskip=0pt\relax}
\providecommand{\BIBentryALTinterwordstretchfactor}{4}
\providecommand{\BIBentryALTinterwordspacing}{\spaceskip=\fontdimen2\font plus
\BIBentryALTinterwordstretchfactor\fontdimen3\font minus
  \fontdimen4\font\relax}
\providecommand{\BIBforeignlanguage}[2]{{%
\expandafter\ifx\csname l@#1\endcsname\relax
\typeout{** WARNING: IEEEtran.bst: No hyphenation pattern has been}%
\typeout{** loaded for the language `#1'. Using the pattern for}%
\typeout{** the default language instead.}%
\else
\language=\csname l@#1\endcsname
\fi
#2}}
\providecommand{\BIBdecl}{\relax}
\BIBdecl

\bibitem{ramanath2005color}
R.~Ramanath, W.~E. Snyder, Y.~Yoo, and M.~S. Drew, ``Color image processing
  pipeline,'' \emph{IEEE Signal Processing Magazine}, vol.~22, no.~1, pp.
  34--43, 2005.

\bibitem{karaimer2016software}
H.~C. Karaimer and M.~S. Brown, ``A software platform for manipulating the
  camera imaging pipeline,'' in \emph{ECCV Proceedings}.\hskip 1em plus 0.5em
  minus 0.4em\relax Springer, 2016, pp. 429--444.

\bibitem{ignatov2020replacing}
A.~Ignatov, L.~Van~Gool, and R.~Timofte, ``Replacing mobile camera {ISP} with a
  single deep learning model,'' in \emph{IEEE/CVF Computer Vision and Pattern
  Recognition Workshops}, 2020.

\bibitem{santos2025isp}
C.~F. G.~d. Santos \emph{et~al.}, ``{ISP} meets deep learning: A survey on deep
  learning methods for image signal processing,'' \emph{ACM Computing Surveys},
  2025.

\bibitem{conde2024nilut}
M.~V. Conde, J.~Vazquez-Corral, M.~S. Brown, and R.~Timofte, ``Nilut:
  Conditional neural implicit 3d lookup tables for image enhancement,'' in
  \emph{Proceedings of the AAAI Conference on Artificial Intelligence},
  vol.~38, no.~2, 2024, pp. 1371--1379.

\bibitem{yoshimura2024pqdynamicisp}
M.~Yoshimura, J.~Otsuka, and T.~Ohashi, ``{PQDynamicISP: }dynamically
  controlled image signal processor for any image sensors pursuing perceptual
  quality,'' \emph{CoRR}, 2024.

\bibitem{nishimura2018automatic}
J.~Nishimura, T.~Gerasimow, R.~Sushma, A.~Sutic, C.-T. Wu, and G.~Michael,
  ``Automatic {ISP} image quality tuning using nonlinear optimization,'' in
  \emph{International Conference on Image Processing (ICIP)}.\hskip 1em plus
  0.5em minus 0.4em\relax IEEE, 2018, pp. 2471--2475.

\bibitem{mosleh2020hardware}
A.~Mosleh, A.~Sharma, E.~Onzon, F.~Mannan, N.~Robidoux, and F.~Heide,
  ``Hardware-in-the-loop end-to-end optimization of camera image processing
  pipelines,'' in \emph{Proceedings of the IEEE/CVF Conference on Computer
  Vision and Pattern Recognition}, 2020.

\bibitem{shi2022refactoring}
Y.~Shi, S.~Li, X.~Jia, and J.~Liu, ``Refactoring {ISP} for high-level vision
  tasks,'' in \emph{International Conference on Robotics and Automation
  (ICRA)}.\hskip 1em plus 0.5em minus 0.4em\relax IEEE, 2022, pp. 2366--2372.

\bibitem{kakarala2022cost}
R.~Kakarala and J.~Wei, ``What is the cost of adding a constraint in linear
  least squares?'' \emph{arXiv preprint arXiv:2201.09935}, 2022.

\bibitem{bianco2013color}
S.~Bianco, A.~R. Bruna, F.~Naccari, and R.~Schettini, ``Color correction
  pipeline optimization for digital cameras,'' \emph{Journal of Electronic
  Imaging}, vol.~22, no.~2, pp. 023\,014--023\,014, 2013.

\bibitem{kim2020dnn}
Y.~Kim, J.~Lee, S.-S. Kim, C.~Yang, T.~Kim, and J.~Yim, ``{DNN}-based {ISP}
  parameter inference algorithm for automatic image quality optimization,''
  \emph{Electronic Imaging}, vol.~32, pp. 1--6, 2020.

\bibitem{tseng2019hyperparameter}
E.~Tseng, F.~Yu, Y.~Yang, F.~Mannan, K.~S. Arnaud, D.~Nowrouzezahrai, J.-F.
  Lalonde, and F.~Heide, ``Hyperparameter optimization in black-box image
  processing using differentiable proxies.'' \emph{ACM Transactions on
  Graphics}, 2019.

\bibitem{yang2020effective}
C.~Yang, J.~Kim, J.~Lee, Y.~Kim, S.-S. Kim, T.~Kim, and J.~Yim, ``Effective
  {ISP} tuning framework based on user preference feedback,'' \emph{Electronic
  Imaging}, vol.~32, pp. 1--5, 2020.

\bibitem{tseng2022neural}
E.~Tseng, Y.~Zhang, L.~Jebe, X.~Zhang, Z.~Xia, Y.~Fan, F.~Heide, and J.~Chen,
  ``Neural {P}hoto-{F}inishing.'' \emph{ACM Transactions on Graphics}, vol.~41,
  no.~6, pp. 238--1, 2022.

\bibitem{kim2023learning}
H.~Kim and K.~M. Lee, ``Learning controllable {ISP} for image enhancement,''
  \emph{IEEE Transactions on Image Processing}, vol.~33, 2023.

\bibitem{wang2024adaptiveisp}
Y.~Wang, T.~Xu, Z.~Fan, T.~Xue, and J.~Gu, ``{AdaptiveISP}: Learning an
  adaptive image signal processor for object detection,'' \emph{Advances in
  Neural Information Processing Systems}, vol.~37, 2024.

\bibitem{yoshimura2023dynamicisp}
M.~Yoshimura, J.~Otsuka, A.~Irie, and T.~Ohashi, ``{DynamicISP}: dynamically
  controlled image signal processor for image recognition,'' in
  \emph{Proceedings of the IEEE/CVF International Conference on Computer
  Vision}, 2023, pp. 12\,866--12\,876.

\bibitem{kim2025ccmnet}
D.~Kim, M.~Afifi, D.~Kim, M.~S. Brown, and S.~J. Kim, ``{CCMNet}: Leveraging
  calibrated color correction matrices for cross-camera color constancy,''
  \emph{arXiv preprint arXiv:2504.07959}, 2025.

\bibitem{shen2017color}
T.~Shen, J.~Wang, T.~Fang, S.~Zhu, and L.~Quan, ``Color correction for
  image-based modeling in the large,'' in \emph{ACCV: Asian Conference on
  Computer Vision}.\hskip 1em plus 0.5em minus 0.4em\relax Springer, 2017, pp.
  392--407.

\bibitem{bonneel2013example}
N.~Bonneel, K.~Sunkavalli, S.~Paris, and H.~Pfister, ``Example-based video
  color grading.'' \emph{ACM Transactions on Graphics}, vol.~32, 2013.

\bibitem{ke2023neural}
Z.~Ke, Y.~Liu, L.~Zhu, N.~Zhao, and R.~W. Lau, ``Neural preset for color style
  transfer,'' in \emph{Proceedings of the IEEE/CVF Conference on Computer
  Vision and Pattern Recognition}, 2023, pp. 14\,173--14\,182.

\bibitem{van2014color}
A.~Van~Hurkman, \emph{Color Correction Look Book: Creative Grading Techniques
  for Film and Video}.\hskip 1em plus 0.5em minus 0.4em\relax Pearson
  Education, 2014.

\bibitem{yamakabe2020tunable}
R.~Yamakabe, Y.~Monno, M.~Tanaka, and M.~Okutomi, ``Tunable color correction
  for noisy images,'' \emph{Journal of Electronic Imaging}, vol.~29, no.~3, pp.
  033\,012--033\,012, 2020.

\bibitem{bychkovsky2011learning}
V.~Bychkovsky, S.~Paris, E.~Chan, and F.~Durand, ``Learning photographic global
  tonal adjustment with a database of input/output image pairs,'' in
  \emph{Proceedings of the IEEE/CVF Conference on Computer Vision and Pattern
  Recognition}.\hskip 1em plus 0.5em minus 0.4em\relax IEEE, 2011, pp. 97--104.

\bibitem{larchenko2025color}
M.~Larchenko, A.~Lobashev, D.~Guskov, and V.~V. Palyulin, ``Color transfer with
  modulated flows,'' in \emph{Proceedings of the AAAI Conference on Artificial
  Intelligence}, vol.~39, no.~4, 2025, pp. 4464--4472.

\bibitem{radford2021learning}
A.~Radford, J.~W. Kim, C.~Hallacy, A.~Ramesh, G.~Goh, S.~Agarwal, G.~Sastry,
  A.~Askell, P.~Mishkin, J.~Clark \emph{et~al.}, ``Learning transferable visual
  models from natural language supervision,'' in \emph{International Conference
  on Machine Learning}.\hskip 1em plus 0.5em minus 0.4em\relax PmLR, 2021.

\bibitem{ilharco_gabriel_2021_5143773}
\BIBentryALTinterwordspacing
G.~Ilharco, M.~Wortsman, R.~Wightman, C.~Gordon \emph{et~al.}, ``{OpenCLIP},''
  2021. [Online]. Available: \url{https://doi.org/10.5281/zenodo.5143773}
\BIBentrySTDinterwordspacing

\bibitem{wang2022exploring}
J.~Wang, K.~C. Chan, and C.~C. Loy, ``Exploring {CLIP} for assessing the look
  and feel of images,'' in \emph{AAAI}, 2023.

\bibitem{wang2022clip}
Z.~Wang, W.~Liu, Q.~He, X.~Wu, and Z.~Yi, ``Clip-gen: Language-free training of
  a text-to-image generator with clip,'' \emph{arXiv preprint
  arXiv:2203.00386}, 2022.

\bibitem{fu2024mgie}
T.-J. Fu, W.~Hu, X.~Du, W.~Y. Wang, Y.~Yang, and Z.~Gan, ``{Guiding
  Instruction-based Image Editing via Multimodal Large Language Models},'' in
  \emph{Int. Conference on Learning Representations}, 2024.

\bibitem{nguyen2024instruction}
T.~T. Nguyen, Z.~Ren, T.~Pham, T.~T. Huynh, P.~L. Nguyen, H.~Yin, and Q.~V.~H.
  Nguyen, ``Instruction-guided editing controls for images and multimedia: A
  survey in {LLM} era,'' \emph{arXiv preprint arXiv:2411.09955}, 2024.

\bibitem{chai2025giftcomesgoldpaper}
\BIBentryALTinterwordspacing
S.~S. Chai, W.~Peng, B.~Hariharan, and H.~Averbuch-Elor, ``Not every gift comes
  in gold paper or with a red ribbon: Exploring color perception in
  text-to-image models,'' 2025. [Online]. Available:
  \url{https://arxiv.org/abs/2508.19791}
\BIBentrySTDinterwordspacing

\bibitem{li2024coco}
Y.~Li, Y.~Bai, S.~Yang, and J.~Liu, ``Coco-lc: Colorfulness controllable
  language-based colorization,'' in \emph{ACM MM}, 2024.

\bibitem{liu2024learnable}
A.~Liu, S.~Mu, and S.~Xu, ``A learnable color correction matrix for {RAW}
  reconstruction,'' \emph{arXiv preprint arXiv:2409.02497}, 2024.

\bibitem{chou2024embedding}
J.~C.-C. Chou and N.~Alam, ``Embedding geometries of contrastive language-image
  pre-training,'' \emph{preprint arXiv:2409.13079}, 2024.

\bibitem{schuhmann2022laionb}
C.~Schuhmann, R.~Beaumont, R.~Vencu \emph{et~al.}, ``{LAION}-5b: An open
  large-scale dataset for training next generation image-text models,'' in
  \emph{Conference on Neural Information Processing Systems}, 2022.

\bibitem{pytorch}
\BIBentryALTinterwordspacing
A.~Paszke, S.~Gross, F.~Massa \emph{et~al.}, ``{PyTorch}: An imperative style,
  high-performance deep learning library,'' \emph{CoRR}, vol. abs/1912.01703,
  2019. [Online]. Available: \url{http://arxiv.org/abs/1912.01703}
\BIBentrySTDinterwordspacing

\bibitem{arias2025color}
G.~Arias, R.~Baldrich, and M.~Vanrell, ``Color in visual-language models:
  {CLIP} deficiencies,'' \emph{preprint arXiv:2502.04470}, 2025.

\bibitem{hasler2003measuring}
D.~Hasler and S.~E. Suesstrunk, ``Measuring colorfulness in natural images,''
  in \emph{Human Vision and Electronic Imaging VIII}, vol. 5007.\hskip 1em plus
  0.5em minus 0.4em\relax SPIE, 2003, pp. 87--95.

\end{thebibliography}

%

\end{document}